\documentclass[10pt]{amsart}

\newtheorem{theorem}{Theorem}[section]
\newtheorem{lemma}[theorem]{Lemma}

\theoremstyle{definition}
\newtheorem{definition}[theorem]{Definition}

\theoremstyle{remark}

\numberwithin{equation}{section}

\newcommand{\e}{{\varepsilon}}

\newcommand{\R}{{\bf R}}
\newcommand{\Z}{{\bf Z}}

\newcommand{\rw}{\rightarrow}

\begin{document}
\baselineskip 13pt

\author{Gregory Eskin}
\address{Department of mathematics, UCLA, Los Angeles, CA 90095-1555, USA}
\author{Hiroshi ISOZAKI}
\address{Institute of Mathematics \\
University of Tsukuba,
Tsukuba, 305-8571, Japan}

\title[Gauge equivalence]{Gauge Equivalence and Inverse Scattering for Long-Range Magnetic Potentials}
\date{December, 24, 2010}

\maketitle

\begin{abstract}
For Schr\"odinger operators with long-range magnetic vector potentials and short range electric scalar potentials in an exterior domain $\Omega$ in ${\bf R}^n$ with $n \geq 2$, we show that there is a one-to-one correspondence between the gauge equivalent classes of Hamiltonians and those of S-matrices, if $\Omega$ is exterior to a bounded convex obstacle.
\end{abstract}

\section{Introduction}

Let $\Omega= \R^n\setminus \overline{\mathcal O}$,  where ${\mathcal O}$  is a bounded domain with smooth boundary, called an {\it obstacle}. We consider a Schr\"odinger operator
\begin{equation}                           \label{eq:1.1}
H=(-\nabla_x-A(x))^2+V(x)
\end{equation}
in $\Omega$ with Dirichlet boundary condition, 
where $A(x)=(A_1,...,A_n)\in C^\infty(\overline \Omega)$  is the magnetic vector potential and $V(x)\in C^\infty(\overline\Omega)$  is the electric scalar potential.  We assume that
$A(x), V(x)$ are real-valued, and satisfy  the following conditions:
$V(x)$  is a short-range potential,  i.e. 
\begin{equation}                             \label{eq:1.2}
|\partial_x^\alpha V(x)|\leq C_\alpha\langle x\rangle^{-1-\e_0-|\alpha|},\ \ \forall\alpha,
\end{equation}
where $\e_0>0$  and  $\langle x\rangle =(1+|x|^2)^{1/2},\ 
A(x)$ is a long-range potential,  i.e.
\begin{equation}                            
|\partial_x^\alpha A(x)|\leq C_\alpha\langle x\rangle^{-1-|\alpha|},\ \ \forall\alpha.
\label{eq:1.3}
\end{equation}
Throughout the paper, $C$'s denote various constants independent of the variable $x$ (and other parameters).
We also assume that $A(x)$ satisfies the {\it transversal gauge condition}
\begin{equation}                                 \label{eq:1.4}
|\partial_x^\alpha(x\cdot A(x))| \leq C_\alpha\langle x\rangle^{-1-|\alpha|},\ \ \forall \alpha.
\end{equation}
It is well-known that the scattering operator $S$  exists  under the conditions (\ref{eq:1.2}), (\ref{eq:1.3}),  (\ref{eq:1.4}) (cf. [LoTh87]). The  properties of $S$ related to the Aharonov-Bohm effect and the  scattering problem were discussed in 
[AB59], [Rui83], [RoYa02], and others.  

In this paper, we shall consider a slightly more restrictive class of long-range magnetic potentials than (\ref{eq:1.3}). Namely, in addition to the conditions (\ref{eq:1.2}), (\ref{eq:1.3}) and (\ref{eq:1.4}), we assume  for some $R > 0$
\begin{equation}                             \label{eq:1.5}
A(x)=A_0(x)+A_1(x) \ \ \mbox{for}\ \ |x|>R,
\end{equation}
where $A_0(x) \in C^\infty({\bf R}^n\setminus\{0\})$,
 homogeneous of degree $-1$, and satisfies
\begin{equation}                           \label{eq:1.6}
x\cdot A_0(x)=0.
\end{equation}
We also assume that $A_1(x)$ is 
a short-range potential decaying like (\ref{eq:1.2}).
The condition (\ref{eq:1.6}) implies 
that $A_1(x)$ satisfies  (\ref{eq:1.4}).

In order to define a gauge transformation group, we need to introduce two spaces of functions. Let $\mathcal L_{\infty,0}$ be the set of real-valued functions $L(x) \in C^{\infty}(\overline{\Omega})$ satisfying
\begin{equation}
|\partial_x^{\alpha}L(x)| \leq C_{\alpha}\langle x\rangle^{-\epsilon_0-|\alpha|}, \quad \forall \alpha,
\label{S1Linfty0}          
\end{equation} 


Let $\mathcal H_0$ be the set of $C^{\infty}({\bf R}^n\setminus\{0\})$-functions,
 homogeneous of degree  0. We sometimes write $\psi(x) \in \mathcal H_0$ as $\psi(\hat x) \in C^{\infty}(S^{n-1})$, where $\hat x = x/|x|$. When $n=2$, we also write $\varphi \in \mathcal H_0$ as $\varphi(\theta)$, where $\theta$  is the polar angle, i.e. $\hat x=(\cos\theta,\sin\theta)$.

Note that when $\varphi(x)\in \mathcal H_0$ the vector  potential 
$\nabla_x \varphi(x)$  satisfies  the  transversal gauge  condition 
\begin{equation}                                  \label{eq:1.8}
x\cdot\nabla_x\varphi(x)=0.
\end{equation}
Indeed,  let $r=|x|$.  
Then   $x\cdot\nabla_x\varphi(x)=r\frac{\partial}{\partial r}\varphi(\hat x)=0$
since  $\varphi(\hat x)$  is  independent  of $r$.


\begin{definition}
Denote by ${\bf G}(\overline \Omega)$ the group of $C^\infty(\overline\Omega)$  functions having the following properties: $|g(x)|=1$  in $\overline\Omega$  and there exists a constant $R > 1$ such that for $|x|>R$
\begin{equation}                             \label{eq:1.7}
g(x)=\left\{
\begin{split}
& e^{i(\psi(\hat x) + L(x))},\quad \psi \in {\mathcal H}_0, \quad L \in \mathcal L_{\infty,0}, \quad \mbox{when}\ \ n\geq 3, \\
& e^{i(m\theta+\varphi(\theta) + L(x))},\quad \varphi \in {\mathcal H}_0,\quad 
 L \in \mathcal L_{\infty,0}, \quad \mbox{when}\ \ n=2,
\end{split}
\right.
\end{equation}
where  $m$ is an integer (which may depend on $g(x)$), and
\begin{equation}                              \label{eq:1.9}
\int_0^{2\pi}\varphi(\theta)d\theta=0.
\end{equation}
We shall call ${\bf G}(\overline\Omega)$  the {\it gauge group}
and the transformation $u(x) \to g(x)u(x)$  the {\it gauge transformation}.
\end{definition}
 
 If $u(x)$ is the solution to
the Schr\"odinger equation $Hu=\lambda u$,
$v(x) = g(x)u(x)$  satisfies $H'v=\lambda v$,  where
$$
H'=\left(-i\frac{\partial}{\partial x}-A'\right)^2+V',
$$
with $V'(x)=V(x)$, and
\begin{equation}                              \label{eq:1.10}
A'(x)=A(x)-ig^{-1}\frac{\partial g}{\partial x}.
\end{equation}
Magnetic potentials $A'(x)$  and $A(x)$  satisfying (\ref{eq:1.10})
are called {\it gauge equivalent}.
Let $\widehat S(\lambda)$  and $\widehat S'(\lambda)$  be the scattering  matrices 
corresponding  to $H$  and $H'$,  respectively.  Then (see,  for example, [Ya06])
\begin{equation}                              
\widehat S'(\lambda)=e^{i\psi(D_x)}\widehat S(\lambda)e^{-i\psi(-D_x)}
\label{S1S=S'ngeq3}
\end{equation}
when $n\geq 3$, and 
\begin{equation}                              
\widehat S'(\lambda)=e^{im\theta(D_x)+i\varphi(D_x)}\widehat S(\lambda)
e^{-im\theta(-D_x)-i\varphi(-D_x)}
\label{S1S'=Sn=2}
\end{equation}
when $n=2$.
Scattering matrices  $\widehat S(\lambda)$ and $\widehat S'(\lambda)$  satisfying (\ref{S1S=S'ngeq3}) or (\ref{S1S'=Sn=2}) are said to be {\it gauge equivalent}. Therefore, if the magnetic vector potentials are gauge equivalent, so are the scattering matrices. A natural question is the validity of the converse assertion. This is an inverse problem. Since the domain $\Omega$ could be multiply-connected, our task shoud be not only the recovery  of the magnetic  field  but the recovery  of the gauge equivalence classes of the electro-magnetic potentials. 

In our previous work \cite{EIO}, we have dealt with the case $n = 2$ (and $\mathcal H_0 \ni \varphi(\theta) = 0$), and proven that if the scattering matrices coincide, the Schr\"odinger operators are gauge equivalent under the equal flux condition. Moreover, outside the convex obstacles, the equal flux condition is satisfied. Therefore, outside the convex obstacle, there is a one-to-one correspondence between the gauge equivalence classes of Schr\"odinger operators and those of S-matrices. The aim of this article is to extend this result to the case $n \geq 2$ including $\mathcal H_0$.

At the first sight, this problem seems to be of the topological nature, and the case with dimension $n \geq 3$ looks simpler than the case $n = 2$. However,
as    it will be shown,  results  for $n=2$  and $n=3$  are more or less the same
and they are determined by the principal part  $A_0(x)$ of the long-range  magnetic
potential.


\section{Inverse scattering problem}
\label{section 2}
The following theorem is one  of our  main results.


\begin{theorem}                           \label{theo:2.1}
Let $n\geq 2$ and $ 
H^{(j)} =\left(-i\frac{\partial}{\partial x}+H^{(j)}(x)\right)^2 +V^{(j)}$
be  two Schr\"odinger operators in $\Omega^{(j)},j=1,2,$ with
 Dirichlet boundary condition on $\partial\Omega^{(j)}$.
Here $A^{(j)}=A_0^{(j)}+A_1^{(j)},j=1,2,$  satisfy (\ref{eq:1.5}), (\ref{eq:1.6}), and $V^{(j)}(x)$ satisfy (\ref{eq:1.2}). Moreover, when $n=2$  we assume that
\begin{equation}                                    \label{eq:2.7}     
|V^{(1)}(x)-V^{(2)}(x)|\leq C_N\langle x\rangle^{-N},\ \ \forall N,
\end{equation}
\begin{equation}                                    \label{eq:2.8}     
|A^{(1)}_1(x)-A^{(2)}_1(x)|\leq C_N\langle x\rangle^{-N},\ \ \forall N.
\end{equation}
 Suppose $\widehat S^{(1)}(\lambda)=
\widehat S^{(2)}(\lambda)$  for all $\lambda>0$  and suppose there exists $R > 0$ such that
$A_0^{(1)}(x)=A_0^{(2)}(x)$  for $|x|>R$.  Then $\Omega^{(1)}=\Omega^{(2)},
\ V^{(1)}=V^{(2)}$  in $\Omega$  and $A^{(1)}$ and $A^{(2)}$ 
are gauge equivalent in $\Omega$.
\end{theorem}

{\bf Proof:}
Let $B_R=\{x: |x|<R\}$  be a ball that contains the complements to  
$\Omega_1$  and $\Omega_2$.  Let $\{x_0+s\omega,s\in\R,\omega\in S^{n-1}\}$
be a line not intersecting $B_R$.  It was shown in 
[Nic00],  [We02], [BaWe07]
  that
$\hat S^{(1)}(\lambda)=
\hat S^{(2)}(\lambda)$  for all $\lambda >0$  implies
\begin{equation}                                 \label{eq:2.1}          
\exp\left(i\int_{-\infty}^\infty A^{(1)}(x_0+s\omega)\cdot \omega ds\right)
 =\exp\left(i\int_{-\infty}^\infty A^{(2)}(x_0+s\omega)\cdot \omega ds\right),
\end{equation}
\begin{equation}                                  \label{eq:2.2}         
\int_{-\infty}^\infty V^{(1)}(x_0+s\omega)ds 
=\int_{-\infty}^\infty V^{(2)}(x_0+s\omega)ds.
\end{equation}
Note that 
(\ref{eq:1.6})  implies  that integrals  in (\ref{eq:2.1})  converge.  Since
we assume  that $A_0^{(1)}=A_0^{(2)}$  we get  from (\ref{eq:2.1})  that
\begin{equation}                                     \label{eq:2.3}      
\exp\left(i\int_{-\infty}^\infty (A_1^{(1)}-A_1^{(2)})(x_0+s\omega)\cdot \omega dt
\right)=1.
\end{equation}
Therefore  $\int_{-\infty}^\infty (A_1^{(1)}-A_1^{(2)})(x_0+s\omega)\cdot \omega dt=2\pi m,$
where  $m$  is integer.  Since this integral  tends to zero  when 
$|x_0|\rightarrow\infty$  we get  that  $m=0$.

Consider first the case $n\geq 3$.  Let $\Pi$ be an arbitrary two-dimensional
plane in the exterior of $B$.
Then it follows from (\ref{eq:2.2}), (\ref{eq:2.3}) (cf. [Nic00], [Es03], [BaWe07])  that
$V^{(1)}=V^{(2)}$  on $\Pi$  and $B_1^{(1)}|_\Pi=B_1^{(2)}|_\Pi$,  where 
$B_1^{(i)}=dA_1^{(i)}$ and $B_1^{(1)}|_\Pi$ is 
the restriction of the two-form $B_1^{(i)}$  to $\Pi$.
Since  $\Pi$ is an arbitrary plane  we get that 
\begin{equation}                                       \label{eq:2.4}     
V^{(1)}=V^{(2)},\ \ B_1^{(1)}=B_1^{(2)} \ \ \mbox{when}\ \ |x|\geq R.
\end{equation}
Therefore there exists $L_1 \in \mathcal L_{\infty,0}$ such that
\begin{equation}                                   \label{eq:2.5}           
A_1^{(2)}-A_1^{(1)}= \nabla_x L_1 \ \ \mbox{for}\ \
|x|\geq R,
\end{equation}
since $\R^n\setminus B_R$  is a simply-connected domain  when $n\geq 3$.  

In the case $n=2$  we get (\ref{eq:2.4}),  (\ref{eq:2.5}) under the assumption 
(\ref{eq:2.7}), (\ref{eq:2.8}) (cf. [He99], [Nic00], [EIO]).

Extend arbitrarily  $L_1(x)$ from $|x|\geq R$  to $\R^n, n\geq 2$.
Make the gauge
transformation  of $H^{(2)}$  with the gauge $g_1(x)=e^{iL_1(x)}$.  Then
$H^{(1)}$ and $H^{(3)}=g_1^{-1}H^{(2)}g_1$ coincide for $|x|\geq R$.  Note that 
$\widehat S^{(3)}(\lambda)=\widehat S^{(2)}(\lambda)$ since 
$\nabla_x L_1$  is a short-range potential.  Now as 
in the proof of Theorem 5.7  in [EIO]  we want to show that 
$\widehat S^{(1)}(\lambda)=\widehat S^{(2)}(\lambda)$ and $H^{(3)}=H^{(1)}$ for
$|x|\geq R$  implies that the Dirichlet-to-Neumann operators $\Lambda^{(1)}(\lambda)$
and $\Lambda^{(3)}(\lambda)$
for $H^{(1)}$ and $H^{(3)}$ respectively coincide on $C=\{|x|=R\}$ for any
$\lambda >0$. 
Theorem  5.7  of [EIO]  is not trivial  in the long-range case.  Fortunately,
the results of \S 2, \S 3,
 4.1, 5.1 of \cite{EIO}  hold for any $n\geq 2$ when  conditions 
(\ref{eq:1.2}), (\ref{eq:1.3}),  (\ref{eq:1.4}) are satisfied.  In particular,  we have the spatial asymptotics of the resolvent and also that of the eigenoperator, whose kernel is the distorted plane wave (cf. Lemma 3.9 and Theorems 4.4  and 4.5  in [EIO]). 
 Next using the BC-method  (cf.  [Be97], [KuLa00], [Es06], [Es07]) 
  we get that $\Lambda^{(1)}(\lambda)=
\Lambda^{(2)}(\lambda)$  for any  $\lambda>0$ implies that $\Omega^{(1)}=\Omega^{(2)},
V^{(1)}=V^{(2)}$  in $\Omega$  and 
$A^{(2)}-A^{(1)}=ig^{-1}\nabla_x g$  in 
$\Omega$ (cf. Theorem 5.8  in [EIO]).
\qed

\medskip
Theorem \ref{theo:2.1} proves the main result under the additional assumption $A_0^{(1)}=A_0^{(2)}$.
In the next two sections we show that this 
condition  is necessary  when  $H^{(1)}$  and  $H^{(2)}$  are 
gauge equivalent  and it       follows from the equality 
of the scattering matrices
$\widehat S^{(1)}(\lambda)=\widehat S^{(2)}(\lambda)$  in the case when 
$\Omega^{(1)}=\Omega^{(2)}$, and it  is an exterior to a convex obstacle.


\section{The two-dimensional case}
\label{section 3}

We first consider the 2-dimensional case. Let us start with the following fact.


\begin{lemma}                               \label{lma:3.1}
Let $A_0(x)\in C^{\infty}({\bf R}^2\setminus\{0\})$, and $A_0(x)$ is homogeneous
of degree $-1$.
Suppose $A_0(x)\cdot x=0$.  Then there exists $a_0(x) \in \mathcal H_0$ such that
\begin{equation}                              \label{eq:3.1}
A_0(x)=\alpha \frac{(-x_2,x_1)}{|x|^2}+ \nabla_xa_0(x),
\end{equation}
\begin{equation}                              \label{eq:3.2}
\alpha=\frac{1}{2\pi}\int_{|x|=R}A_0(x)\cdot dx,\ \ \int_{0}^{2\pi}a_0(\cos\theta,\sin\theta)d\theta=0.
\end{equation} 
\end{lemma} 

{\bf Proof:}
Let $A_0(x)=(A_{01}(x),A_{02}(x))$.  Then 
$A_0(x)\cdot x=0$ implies $x_1A_{01}+x_2A_{02}=0$, and
$$
\frac{A_{02}}{x_1}=\frac{A_{01}}{-x_2}=\frac{a(x)}{|x|^2},
$$
where $a(x)$  is homogeneous of degree zero.
Therefore
\begin{equation}                                  \label{eq:3.3}
A_{01}=-\frac{x_2}{|x|^2}a(x),\ \ A_{02}=\frac{x_1}{|x|^2}a(x).
\end{equation}
Since $a(x)=a(x/|x|)$  we have in polar coordinates
$x_1=r\cos\theta,x_2=r\sin\theta$  that $a(x/|x|)=a(\cos\theta,\sin\theta)$.
Let $\hat a(\theta)=a(\cos\theta,\sin\theta)$, i.e. $\hat a(\theta)$ is $2\pi$-periodic function of $\theta$,  and let
\begin{equation}                                     \label{eq:3.4}
\alpha=\frac{1}{2\pi}\int_0^{2\pi}\hat a(\theta)d\theta.
\end{equation}
Note that $\alpha$ in (\ref{eq:3.2}), (\ref{eq:3.4}) is the same. 
 Moreover, $\alpha$ can be defined as 
the limit: $\alpha=\lim_{R\rw\infty}\frac{1}{2\pi}\int_{|x|=R}A(x)\cdot dx$.

Let $b(\theta)=\hat a(\theta)-\alpha$.  Then  $\int_0^{2\pi}b(\theta)d\theta=0$.                    
Therefore $b(\theta)$ can be represented in the form
$$
b(\theta)=\frac{\partial}{\partial\theta}\hat a_0(\theta),
$$
where $\hat a_0(\theta)$ is $2\pi$-periodic 
and 
$$
\int_0^{2\pi}\hat a_0(\theta)d\theta=0.
$$
Let $a_0(x/|x|) \in \mathcal H_0$ be such   that $a_0(\cos\theta,\sin\theta)=\hat a_0(\theta),\ 
\theta=\tan^{-1}(x_1/x_2)$.
We have 
$$
\frac{\partial a_0}{\partial x_1}=\frac{\partial\hat a_0(\theta)}{\partial\theta}
\frac{\partial\theta}{\partial x_1}
=\frac{x_2}{|x|^2}\frac{\partial \hat a_0(\theta)}{\partial\theta},
$$
$$
\frac{\partial a_0}{\partial x_2}=\frac{\partial\hat a_0(\theta)}{\partial\theta}
\frac{\partial\theta}{\partial x_2}
=-\frac{x_1}{|x|^2}\frac{\partial \hat a_0(\theta)}{\partial\theta}.
$$
Therefore we have proven
\begin{equation}
\begin{split}                                       \label{eq:3.5}
A_0(x) & =\frac{(-x_2,x_1)}{|x|^2}\hat a(\theta)=\frac{(-x_2,x_1)}{|x|^2}\left(\alpha+
\frac{\partial \hat a_0(\theta)}{\partial\theta}\right)\\
& =\alpha\frac{(-x_2,x_1)}{|x|^2}+\nabla_xa_0(x). 
\ \ \ \ \ \ \ \ \ \ \ \ \ \ \ \ \ \ \ \ \ \ \ \ \qed
\end{split}
\end{equation}

\medskip
It follows from Lemma \ref{lma:3.1} that when $n=2$  any magnetic potential of the form  
(\ref{eq:1.5}),  (\ref{eq:1.6})  
is gauge equivalent  in $\Omega$  to the Aharonov-Bohm  type potential
\begin{equation}                                   \label{eq:3.6}
A_\alpha=\alpha\frac{(-x_2,x_1)}{|x|^2}+A_1'(x),
\end{equation}
where $A_1'$  is a short-range  potential.

It was shown in  [Ya06], [RoYa02]  
that the kernel of scattering matrix $\widehat S_\alpha(\lambda)$
corresponding to (\ref{eq:3.6})  has the form 
\begin{equation}                                    \label{eq:3.7}
S_\alpha(\theta,\theta')=S_\alpha(\theta-\theta')+S_\alpha'(\theta,\theta'),
\end{equation}
where
\begin{equation}                                 \label{eq:3.8}
S_\alpha(\theta)=\cos(\alpha\pi)\delta(\theta)
+i\frac{\sin(\alpha\pi)}{\pi}{\rm p.v.} \frac{e^{i[\alpha]\theta}}{1-e^{i\theta}},
\end{equation}
\begin{equation}                                  \label{eq:3.9}
|S_\alpha'(\theta,\theta')|\leq  C|\theta-\theta'|^{-\delta},\ 0<\delta<1,
\end{equation}
$[\alpha]$  is the integer part of $\alpha$.  (In [Ya06], [RoYa02]  the case $\Omega=\R^2$
was considered.  The extension to the case  
of domains with obstacles was shown in [EIO]).

It follows from Lemma \ref{lma:3.1}  and (\ref{S1S'=Sn=2}) that the kernel of the 
scattering matrix $\widehat S(\lambda)$ 
corresponding 
to $A(x)=A_0+A_1$  has the form (cf.  [RoYa02], [Ya06])
\begin{equation}                              \label{eq:3.10}
S(\theta,\theta')=e^{i\hat a_0(\theta)-i\hat a_0(\pi+\theta')}(S_\alpha(\theta-\theta')
+S_\alpha'(\theta,\theta')).
\end{equation}

The next theorem shows that in the case of a convex obstacle 
$\hat S^{(1)}(\lambda)=\hat S^{(2)}(\lambda)$
implies that $A_0^{(1)}=A_0^{(2)}$ if  $\alpha$  is not an integer.


\begin{theorem}                                      \label{theo:3.2}
Let $H^{(j)}=\left(-i\frac{\partial}{\partial x}+A^{(j)}\right)^2+V^{(j)}$ be two
Schr\"odinger operators in the same domain 
$\Omega\subset\R^2$, where $\Omega$ is the exterior to a convex obstacle.  Suppose $V^{(j)},j=1,2,$  are short-range 
potentials and 
$$
A^{(j)}=\alpha_j\frac{(-x_2,x_1)}{|x|^2}+\nabla_x \varphi_j(x)+A_1^{(j)},\quad j=1,2,
$$ 
where $\varphi_j(x) \in \mathcal H_0$,$\ \int_{S^1}\varphi_j(\hat x)d\sigma=0$, 
\ $A_1^{(j)}$  are short-range potentials, $j=1,2.$
Suppose $\widehat S^{(1)}(\lambda)=\widehat S^{(2)}(\lambda)$ for all $\lambda > 0$ and $\alpha_1$  is not 
integer.  Suppose also that (\ref{eq:2.7}), (\ref{eq:2.8}) hold.  Then
\begin{equation}                                     \label{eq:3.11}
\alpha_1=\alpha_2,\ \ \varphi_1(x)=\varphi_2(x),
\end{equation}
$V^{(1)}=V^{(2)}$  in $\Omega$  and there exists  $L_1(x) \in \mathcal L_{\infty,0}$ 
such that $A_1^{(2)}-A_1^{(1)}=\nabla_x L_1$.
\end{theorem}

{\bf Proof:}  As has been proved in [Nic00], [We02], $\widehat S^{(1)}(\lambda)=\widehat S^{(2)}(\lambda)$  
for all $\lambda>0$  implies  (\ref{eq:2.1}),  (\ref{eq:2.2})  for all lines
$x=x_0+s\omega$  not intersecting 
$\Omega$.  Let $\omega=(\cos\theta,\sin\theta),\ x_1=x_{10}+s\cos\theta,\ 
x_2=x_{20}+s\sin\theta, \ x_0\cdot\omega=0$.  Then
\begin{equation}                                       \label{eq:3.12}
\int_{-\infty}^\infty\alpha\frac{(-x_2,x_1)}{|x|^2}\cdot\omega ds
=\alpha \int_{-\infty}^\infty\frac{|x_0|}{|x_0|^2+s^2}ds=\alpha\pi,
\end{equation}
\begin{eqnarray}                                     \label{eq:3.13}
\int_{-\infty}^\infty\frac{\partial\varphi(x_0+s\omega)}{\partial x}\cdot\omega ds=
\int_{-\infty}^\infty\frac{d}{ds}\varphi(x_0+s\omega)ds 
\\
\nonumber
=\varphi(x_0+s\omega)|_{-\infty}^\infty
=\varphi(\frac{1}{|s|}x_0+\frac{s}{|s|}\omega)|_{-\infty}^\infty=\varphi(\omega)-\varphi(-\omega),
\end{eqnarray}
where we used that
$\varphi$  is homogeneous of degree  0.
Note that $\int_{-\infty}^\infty A_1^{(j)}(x_0+s\omega)\cdot \omega ds\rw 0$  
when $|x_0|\rw \infty, \ x_0\cdot\omega=0$.  Therefore taking the limit 
in (\ref{eq:2.1})  when  $|x_0|\rw \infty$  we get
\begin{equation}                                \label{eq:3.14}
(\alpha_2-\alpha_1)\pi+\varphi(\omega)-\varphi(-\omega)=2\pi m,
\end{equation}
for some integer $m$,
where $\varphi(\omega)=\varphi_2(\omega)-\varphi_1(\omega)$.  It follows from 
(\ref{eq:3.14})  that $\varphi(\omega)-\varphi(-\omega)$ does not
depend  on $\omega$.   Since  $\varphi(\omega)-\varphi(-\omega)$  is odd,
there exists $\omega_0$ such that $\varphi(\omega_0)-\varphi(-\omega_0)=0.$
Therefore
\begin{equation}                                \label{eq:3.15} 
\varphi(\omega)-\varphi(-\omega)=0,\ \ \forall\omega\in S^1,
\end{equation}
\begin{equation}                                 \label{eq:3.16}
\alpha_2-\alpha_1=2m.
\end{equation}
It follows from (\ref{eq:3.14})  and (\ref{eq:2.1})
that (\ref{eq:2.3}) holds.
Therefore using (\ref{eq:2.2}), (\ref{eq:2.3}),  (\ref{eq:2.7}) and (\ref{eq:2.8}), we get
as in the proof of Theorem  \ref{theo:2.1} (cf. [Nic00], [BaWe07], [Es03])  that
$V^{(1)}=V^{(2)}$  in $\Omega$  and  (\ref{eq:2.5})  holds  in
$\Omega$.   As in the proof of Theorem \ref{theo:2.1},  let  
$H^{(3)}=g_1^{-1}H^{(2)}g_1$  where  $g_1=e^{iL_1}$.   Let
$\widehat S^{(3)}(\lambda)$  is   the scattering 
matrix corresponding to $H^{(3)}$.  Note also that $H^{(3)}$  has the form
$$
H^{(3)}=\left(-i\frac{\partial}{\partial x}+\alpha_2\frac{(-x_2,x_1)}{|x|^2}+
\nabla_x\varphi_2 +A_1^{(1)}\right)^2+V^{(1)}(x).
$$
Let  $g(x)=e^{i2m\theta +i\varphi(\hat x)}$ 
where  $\hat x=(\cos\theta,\sin\theta)$.   Then
$g^{-1}H^{(3)}g=H^{(1)}$.  Therefore (cf.  (\ref{S1S'=Sn=2}))
\begin{equation}                           \label{eq:3.17}
S^{(3)}(\theta,\theta')=e^{i2m(\theta-\theta')+i\varphi(\omega)-i\varphi(-\omega)}S^{(1)}(\theta,\theta'),
\end{equation}
where (cf. (\ref{eq:3.7}),  (\ref{eq:3.8}))
\begin{equation}                            \label{eq:3.18}
S^{(1)}(\theta,\theta')=(S_{\alpha_1}(\theta-\theta')
+S'_{\alpha_1}(\theta,\theta'))e^{i\varphi_1(\omega)-i\varphi_1(-\omega)},
\end{equation}
\begin{equation}                              \label{eq:3.19}
|S'_{\alpha_1}(\theta,\theta')|\leq C|\theta-\theta'|^{-\delta},\ \ 0\leq \delta<1.
\end{equation}
Since $\widehat S^{(1)}=\widehat S^{(3)}$  we get
\begin{equation}                              \label{eq:3.20}
(e^{i2m(\theta-\theta')+i(\varphi(\omega)-\varphi(-\omega))}-1)S^{(1)}(\theta,\theta')=0.
\end{equation}
Denote $\varphi_0(\theta)=\varphi(\cos\theta,\sin\theta)$.  Then 
$\varphi(-\cos\theta,-\sin\theta)=
\varphi(\cos(\theta+\pi),\sin(\theta+\pi))=\varphi_0(\theta+\pi)$. We have
$\varphi_0(\theta)-\varphi_0(\theta+\pi)=0,\ \forall\theta$.
Using the Taylor formula we get
\begin{equation}
\begin{split}                                  
\varphi_0(\theta)-\varphi_0(\theta'+\pi)& =\varphi_0(\theta')+\varphi'_0(\theta')(\theta-\theta') +O((\theta-\theta')^2)-\varphi_0(\theta'+\pi)
\\
& =\varphi'_0(\theta')(\theta-\theta')+O((\theta-\theta')^2).
\end{split}
\label{eq:3.21}
\end{equation}
Since $\alpha_1$  is not an integer  we have for arbitrary $\theta_0'$:
\begin{equation}                                    \label{eq:3.22}
|S^{(1)}(\theta,\theta')|=|S_{\alpha_1}(\theta-\theta')+S'_{\alpha_1}(\theta,\theta')|
\geq C|\theta-\theta_0'|^{-1}
\end{equation}
for all $0<\theta-\theta'<\e, \ \e$  is small.  Take $\theta_0'$  such that
$\varphi'_0(\theta_0')=0$.
Then using (\ref{eq:3.21}) and assuming $m\neq0$ we get
\begin{equation}                                               \label{eq:3.23}
|e^{i2m (\theta-\theta_0')+i(\varphi_0(\theta_0')-\varphi_0(\theta_0'+\pi))}-1|
\geq C|\theta-\theta_0'|.
\end{equation}
Therefore the left-hand side of (\ref{eq:3.20})  is nonzero for $0<\theta-\theta_0'<\e$.  This
contradiction  
proves that 
$2m=\alpha_2-\alpha_1=0$.  

Assuming  that $\varphi_0(\theta)$  is not zero
we can find $\theta_0''$  such that $\varphi'(\theta_0'')\neq 0$.  Then taking into account
that $m=0$ we get
\begin{equation}                                         \label{eq:3.24}
|e^{i(\varphi_0(\theta)-\varphi_0(\theta_0''+\pi))}-1|\geq C|\theta-\theta_0''|
\ \ \mbox{for}\ \ |\theta-\theta_0''|\ \ \mbox{small}.
\end{equation}
Again (\ref{eq:3.24})  and (\ref{eq:3.22})  contradict (\ref{eq:3.20}).
Therefore  
$\varphi'_0(\theta)=0$  for all $\theta$  and hence  $\varphi_0(\theta)=\varphi_2(\theta) - \varphi_1(\theta)=0,\ \forall \theta$.
\qed 

\bigskip
The following theorem shows that the condition $A_0^{(1)}=A_0^{(2)}$  is
necessary if one assumes that $\widehat S^{(1)}(\lambda)=\widehat S^{(2)}(\lambda)$  and $H^{(1)}$  and $H^{(2)}$   are  gauge  equivalent.


\begin{theorem}                                     \label{theo:3.3}
Let $\Omega$ be an arbitrary domain with obstacles.  
Let $H^{(j)}=(-i\frac{\partial}{\partial x}+A^{(j)}(x))^2+V^{(j)}(x)$ be the Schr\"odinger 
 operators in $\Omega$
as in Theorem \ref{theo:2.1}.
Suppose 
$\widehat S^{(1)}(\lambda)=\widehat S^{(2)}(\lambda)$ and suppose  that $H^{(1)}$ and 
$H^{(2)}$ are gauge equivalent,
i.e. there exists $g(x) \in {\bf G}(\overline{\Omega})$  such that
$H^{(1)}=g^{-1}H^{(2)}g$.  Suppose  also that
$$
\alpha_1=\lim_{R\rw \infty}\frac{1}{2\pi}\int_{|x|=R}A^{(1)}(x)\cdot dx
$$
is not an integer.  Then $A_0^{(1)}(x)=A_0^{(2)}(x).$
\end{theorem}

{\bf Proof:}  Denote $H^{(3)}=g^{-1}H^{(2)}g$.  Then $H^{(3)}=H^{(1)}$  and therefore
$\widehat S^{(3)}(\lambda)=\widehat S^{(1)}(\lambda)$.  It follows  from (\ref{S1S'=Sn=2})
that 
$$
S^{(3)}(\theta,\theta')=
e^{im(\theta-\theta')+i(\varphi_0(\theta)
-\varphi_0(\theta+\pi))}S^{(2)}(\theta,\theta').
$$
Since  $S^{(2)}(\theta,\theta')=S^{(1)}(\theta,\theta')$  we get  the same equation  as in 
(\ref{eq:3.20}):

\begin{equation}                              \label{eq:3.25}
(e^{i2m(\theta-\theta')+i(\varphi_0(\theta)-\varphi_0(-\theta+\pi))}-1)
S^{(1)}(\theta,\theta')=0,
\end{equation}
i.e.  we are in the same situation as in the proof  of Theorem  \ref{theo:3.2}.
Therefore $m=0,\varphi_0(\theta)=0$.  Hence $A_0^{(2)}=A_0^{(1)}$.
\qed

\bigskip
On the basis of Theorem \ref{theo:3.2}  
we can  formulate our ultimate result in 2-dimensions.


\begin{theorem}                                       \label{theo:3.4}
Let  $\Omega$  be  the exterior of a convex obstacle. Let  $H^{(j)},j=1,2,$  be the
same as in Theorem \ref{theo:3.2}.  Let 
$\widehat S^{(j)}(\lambda)$  be the scattering matrices corresponding to $H^{(j)}, j=1,2.$ If
$\widehat S^{(1)}(\lambda)$ and $\widehat S^{(2)}(\lambda)$
are gauge equivalent for all $\lambda > 0$, then $V^{(1)}=V^{(2)}$ and  $A^{(1)}$ and $A^{(2)}$  are
gauge equivalent.
\end{theorem}

{\bf Proof:}
Since 
$\widehat S^{(1)}(\lambda)$ and $\widehat S^{(2)}(\lambda)$
are gauge equivalent there exists $g(x)=e^{im\theta+i\varphi_0(\theta)}$  such 
that (\ref{S1S'=Sn=2})  holds. Let $H^{(3)}=g^{-1}H^{(2)}g$.  Then 
$\widehat S^{(1)}(\lambda)=\widehat S^{(3)}(\lambda)$
where 
$\widehat S^{(3)}(\lambda)$ 
is the scattering matrix for $H^{(3)}$. 
 Also we have that $A_0^{(3)}$ is gauge  equivalent to  $A_0^{(2)}$, $A_1^{(3)}=A_1^{(2)}$,
$V^{(3)}=V^{(2)}$.
Since 
$\widehat S^{(3)}(\lambda)=\widehat S^{(1)}(\lambda)$
we get applying Theorem \ref{theo:3.2}  that
$A_0^{(3)}=A_0^{(1)}$, $V^{(3)}=V^{(1)}$  and 
$A_1^{(3)}-A_1^{(1)}=\nabla_xL_1$.
Therefore
$A^{(2)}=A_0^{(2)}+A_1^{(2)}$
is gauge equivalent  to
$A^{(1)}=A_0^{(1)}+A_1^{(1)}$  and $V^{(1)}=V^{(2)}$.
\qed

\bigskip
We end this section with a conjecture that Theorem \ref{theo:3.4} holds for 
arbitrary domain with obstacles.


\section{The case $n\geq 3$}
\label{section 4}

The following theorem  is the analog of Theorem  \ref{theo:3.2} for $n\geq 3$.


\begin{theorem}                                    \label{theo:4.1}
Let $\Omega$  be  the exterior of a convex obstacles in $\R^n,n\geq 3$.
Let $H^{(j)}=(-i\frac{\partial}{\partial x}-A^{(j)})^2+V^{(j)}(x)$ be  two          
Schr\"odinger  operators as in 
Theorem \ref{theo:2.1}.
Suppose 
${\rm curl}\,A_0^{(1)}\neq 0$, and  that
$\widehat S^{(1)}(\lambda)=\widehat S^{(2)}(\lambda)$  for all $\lambda>0$.
Then
$V^{(1)}=V^{(2)}$, $A_1^{(2)}-A_1^{(1)}=\nabla_xL_1$ with $L_1\in \mathcal L_{\infty,0}$ and
$A_0^{(1)}=A_0^{(2)}$.
\end{theorem}

{\bf Proof:}
Consider (\ref{eq:2.1}), (\ref{eq:2.2}).  Let $A_0=
A_0^{(2)}-A_0^{(1)}$,  $A_1=A_1^{(2)}-A_1^{(1)}$.  It follows from  (2.3)
$\int_{-\infty}^\infty(A_0+A_1)(x_0+s\omega)\cdot \omega ds=2\pi m$
where  $m\in\Z$.   As  in Theorem  \ref{theo:2.1}  take any  two-dimensional  plane 
$\Pi$  in $\Omega$.  The  proof  in [Es03],  pp 990-991,  give  that 
$B|_\Pi=0$  even  when  $m\neq 0$.   Here  $B=B_0+B_1, B_0=dA_0, B_1=dA_1,  B$  is
the restriction  of 2-form $B$  to $\Pi$.
Since  $\Pi$  is arbitrary  we have as in Theorem \ref{theo:2.1}
 that $B_0+B_1=0,  
\ V^{(1)}=V^{(2)}$  in $\Omega$. Note that   
$B_0(x)=b(x)/|x|^2$ where $b(x)$ is homogeneous of degree  zero 
 and $|B_1(x)|\leq C\langle x\rangle^{-2-\e_0}$.
Multiplying  $B_0+B_1=0$  by  $|x|^2$  and  taking the limit when  $|x|\rw \infty$
we get that $b(x)=0$.  Therefore  $B_0=0$  and $B_1=0$.  Since $\Omega$  is
a simply connected domain,  we have         
$A_1=\nabla_x L_1,\ A_0=\nabla_x\varphi$,  where
$L_1\in \mathcal L_{\infty,0}$  and $\varphi(x) \in \mathcal H_0$.   Since
 $\int_{-\infty}^\infty\nabla_xL_1\cdot \omega ds=0$,
we get  from (\ref{eq:2.1})  that $\int_{-\infty}^\infty A_0\cdot\omega ds=2\pi m$.  

Therefore  as in (\ref{eq:3.13})
\begin{equation}                                          \label{eq:4.1}
2\pi m=\int_{-\infty}^\infty\frac{\partial \varphi}{\partial x}(x_0+s\omega)\cdot\omega ds
=\int_{-\infty}^\infty
\frac{d}{ds}\varphi(x_0+s\omega)ds =\varphi(\omega)-\varphi(-\omega).
\end{equation}
Changing  $\omega$  to $-\omega$  we get  $m=0$,  i.e.  
$\varphi(\omega)-\varphi(-\omega)=0, \forall\omega\in S^{n-1}$.
Let  $g(x)=e^{i\varphi(\omega)+iL_1(x)}$  and let $H^{(3)}=g^{-1}H^{(2)}g$.
Then  $H^{(3)}=H^{(1)}$  and therefore $\widehat S^{(3)}=\widehat S^{(1)}$.
By (\ref{S1S=S'ngeq3}) 
we have $\widehat S^{(3)}(\omega,\omega')=
\widehat S^{(2)}(\omega,\omega')e^{i\varphi(\omega)-i\varphi(-\omega')}$.
Since $\widehat S^{(1)}=\widehat S^{(2)}$  we have
\begin{equation}                                 \label{eq:4.2}
(e^{i(\varphi(\omega)-\varphi(-\omega'))}-1)S^{(1)}(\omega,\omega')=0.
\end{equation}
As in the proof  of Theorem \ref{theo:3.2}, we have by the Taylor formula
\begin{equation}                                  \label{eq:4.3}
e^{i(\varphi(\omega)-\varphi(-\omega'))}=e^{i(\varphi_x(\omega')\cdot (\omega-\omega')+
O(|\omega-\omega'|^2)},
\end{equation}
since $\varphi(\omega')-\varphi(-\omega')=0$.
Therefore  if $\omega_0'$ is such  that $\varphi_x(\omega_0')\neq 0$ then
\begin{equation}                                 \label{eq:4.4}
|e^{i(\varphi(\omega)-\varphi(-\omega')}-1|\geq C|\omega-\omega'|,
\end{equation}
when $|\omega-\omega'|$  is small,   $\omega'$  is close to $\omega_0'$
and $(\omega-\omega')\cdot\varphi_x(\omega_0')\neq 0$.
Since  $\mbox{curl\ } A_0^{(1)}\neq  0$,  we
 get  that $\nabla_x I(x,\omega)\neq 0$, 
where $I(x,\omega)=\int_{-\infty}^\infty A_0^{(1)}(x+s\omega)\cdot \omega ds$.
  Then it follows from [Ya06] 
that $S^{(1)}(\omega,\omega')$  contains a principal  value 
integral  and therefore $S^{(1)}(\omega,\omega')\neq 0$  when (\ref{eq:4.4})  holds.
This contradicts (\ref{eq:4.2}). Therefore we must have  $\varphi =0$, i.e. 
$A_0^{(1)}=A_0^{(2)}$.
\qed

\bigskip

The proof  of the following Theorems  are the same  
as that of Theorems \ref{theo:3.3}  and \ref{theo:3.4}  with the reference to Theorem \ref{theo:3.2} replaced by the reference to Theorem \ref{theo:4.1}.


\begin{theorem}                                \label{theo:4.2}
Suppose  $\Omega_1=\Omega_2=\Omega$  is an  arbitrary 
domain  with
obstacles in $\R^n,n\geq 3$,  and 
$H^{(1)}$  and $H^{(2)}$ are  the same as in Theorem 4.1.
Suppose $\mbox{curl\ }A_0^{(1)}\not\equiv 0$.  If 
$\widehat S^{(1)}(\lambda)=\widehat S^{(2)}(\lambda)$
and if $H^{(1)}$  and $H^{(2)}$ are  gauge equivalent,  then  $A_0^{(1)}=A_0^{(2)}$.
\end{theorem}


\begin{theorem}                                \label{theo:4.3}
Suppose  $\Omega_1=\Omega_2=\Omega \subset \R^n, n\geq 3,$  
where $\Omega$  is  the exterior  of a convex obstacle. Let $H^{(j)} = \left(-\frac{\partial}{\partial x} + A^{(j)}\right)^2 + V^{(j)}$,
$\widehat S^{(j)}(\lambda),\ j=1,2,$  be  the same as in Theorem \ref{theo:4.1}.
If $\widehat S^{(1)}(\lambda)$  and $\widehat S^{(2)}(\lambda)$ are gauge equivalent for all $\lambda > 0$, then
 $H^{(1)}$  and $H^{(2)}$ are  gauge equivalent.
\end{theorem}
It follows from  Theorem  \ref{theo:4.2}  that the assertion:  
$\widehat S^{(1)}(\lambda)=\widehat S^{(2)}(\lambda)$  implies 
$A_0^{(1)}=A_0^{(2)}$  is necessary  when $H^{(1)}$  and  $H^{(2)}$  are  gauge  equivalent.
This leads  us  to a conjecture that Theorem \ref{theo:4.1} and consequently Theorem 
\ref{theo:4.3}  are true  for any obstacle.

\end{document}